\documentclass{appolb}

\usepackage{graphicx}
\usepackage{amsmath}

\begin{document}
\title{Asymptotic behavior of the Speed of Sound in Dense Matter 
\thanks{Presented at V4-HEP3, Prague, Czech Republic, October 2024}%
}
\author{Udita Shukla, Pok Man Lo
\address{
Institute of Theoretical Physics,\\ 
University of Wroc\l aw, Wroc\l aw, Poland
}
\\[3mm]
}
\maketitle
\begin{abstract}
We show that a class of NJL-like models fails to reproduce the expected conformal limit of the speed of sound, making them unsuitable for analyzing the equation of state of dense matter. We then demonstrate how this issue can be resolved within a simple dynamical quark model.
\end{abstract}
  
\section{Introduction}

When a thermodynamic variable, such as the baryochemical potential, becomes
large compared to all other intrinsic scales (masses, couplings), it defines the
dominant energy scale in the system. In this limit, physical observables must
scale with this parameter according to dimensional analysis. This is the key feature of conformal behavior.

This conformal behavior may not emerge if an interaction scale—whether intrinsic to the model or dynamically generated—becomes dominant and grows faster than the controlling variable. In such cases, the system scales with this new interaction scale instead. Fortunately, QCD features asymptotic freedom: the interaction strength decreases at high energies, and we expect that the conformal limit of quarks and gluons is eventually recovered.

Modeling QCD at low and intermediate energies is a complex
task~\cite{Baym:2017whm}. A common strategy is to use effective models that
incorporate spontaneous chiral symmetry breaking, such as the Nambu–Jona-Lasinio
(NJL) model~\cite{Klevansky1992,Buballa:2003qv}. However, in its simplest form, the NJL model lacks the correct asymptotic behavior and, as we will show, fails to describe conformal matter at high baryon densities.

\section{Failure of NJL-like Models with Local Interactions}

The standard NJL model describes quasi-quark dressing via effective mass and chemical potential, governed by the gap equations:
\begin{equation}
    \begin{split}
        M &= m - 2 G_S n_S \\
        \mu^\prime &= \mu - 2 G_V n_V,
    \end{split}
    \label{eq:gap01}
\end{equation}
where $n_S$ and $n_V$ are scalar and vector densities given by
\begin{equation}
    \begin{split}
        n_S &= N_c N_f \int \frac{d^3 q}{(2 \pi)^3} \frac{-4 M}{2 E_q} \left( 1 - N_{\rm th}(E_q) - \bar{N}_{\rm th}(E_q) \right) \\
        n_V &= N_c N_f \int \frac{d^3 q}{(2 \pi)^3} \frac{4}{2} \left( N_{\rm th}(E_q) - \bar{N}_{\rm th}(E_q) \right),
    \end{split}
\end{equation}
with
\begin{equation}
    \begin{split}
        E_q &= \sqrt{q^2 + M^2} \\
        N_{\rm th}(E) &= \frac{1}{e^{\beta (E - \mu^\prime)} + 1}, \quad
        \bar{N}_{\rm th}(E) = \frac{1}{e^{\beta (E + \mu^\prime)} + 1}.
    \end{split}
\end{equation}
In the $T \to 0$ limit, thermal distributions reduce to step functions:
\begin{equation}
    N_{\rm th}(E_q) \rightarrow \theta(\mu^\prime - E_q),
\end{equation}
with antiparticle contributions vanishing. At large density, the medium part of $n_S$ cancels the vacuum term:
\begin{equation}
    \begin{split}
        n_S &\rightarrow 0 \\
        n_V &\rightarrow N_c N_f \frac{p_F^3}{3 \pi^2}, \quad
        p_F = \sqrt{{\mu^\prime}^2 - M^2}.
    \end{split}
\end{equation}

In the chiral limit ($m = 0$), as $\mu \to \infty$ and $M \to 0$, the dressed chemical potential satisfies:
\begin{equation}
    \mu^\prime = \mu - N_c N_f \frac{2 G_V}{3 \pi^2} {\mu^\prime}^3,
\end{equation}
leading to the asymptotic scaling:
\begin{equation}
    \mu^\prime \propto \mu^{1/3} G_V^{-1/3}.
\end{equation}
Hence, the density follows:
\begin{equation}
    n_V \rightarrow \frac{\mu}{2 G_V} \propto {\mu^\prime}^3,
\end{equation}
and the speed of sound at $T=0$ becomes:
\begin{equation}
    c_s^2 = \frac{\partial P}{\partial \epsilon} \approx \frac{\partial \ln
    \mu}{\partial \ln n_V} \rightarrow 1,
\end{equation}
violating the expected conformal limit $c_s^2 = \frac{1}{3}$.

This issue has affected many models of dense equations of state (EoSs), 
often prompting the introduction of ad hoc elements such as form factors or density-dependent couplings~\cite{DB01,Ivanytskyi:2024zip}. 
While these adjustments may offer practical solutions, they lack a solid theoretical foundation and do not provide deeper physical insight. 
In contrast, we demonstrate that models inspired by QCD with asymptotically free interactions offer a more principled approach, leading to a resolution that is both natural and consistent with the expected high-density behavior.

\begin{figure}
	\resizebox{0.9\textwidth}{!}{%
	\includegraphics{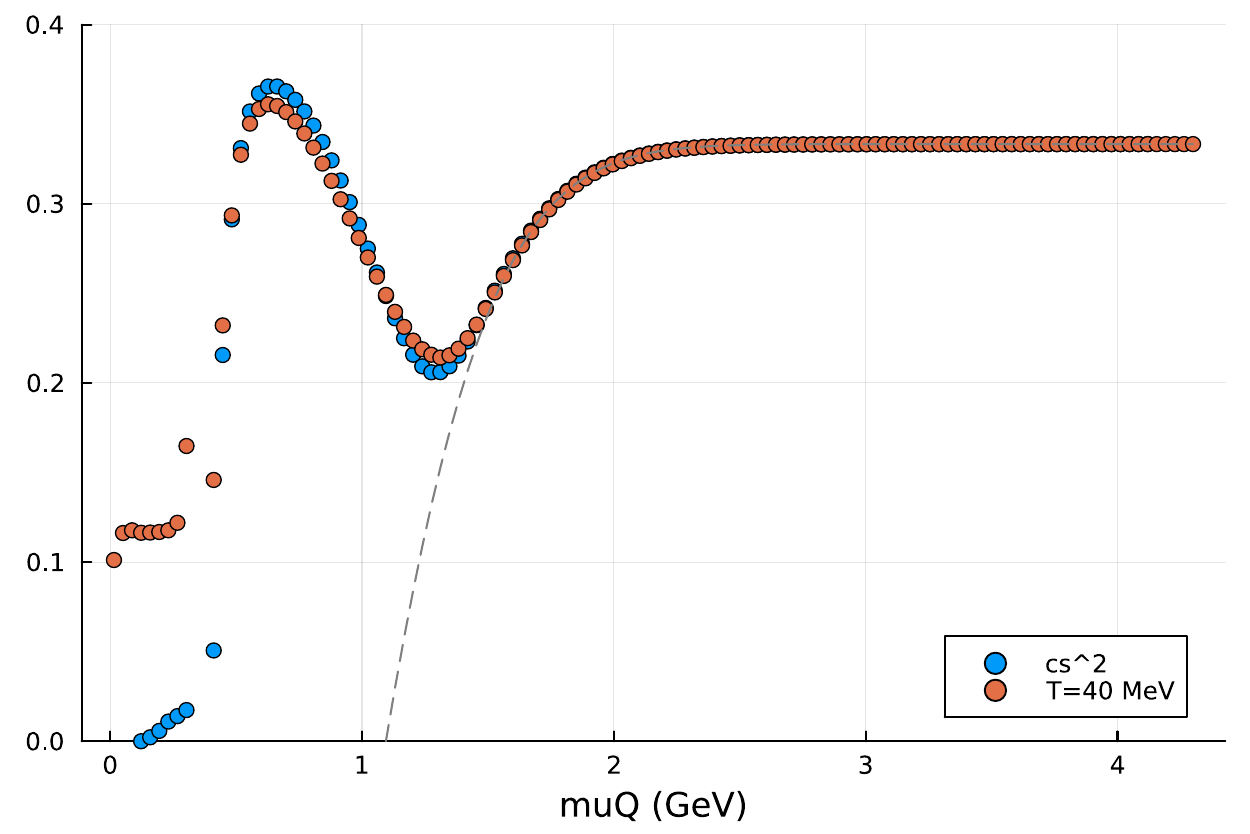}
	}
    \caption{Speed of sound computed in a separable model, compared to the asymptotic formula.}
    \label{fig1}
\end{figure}

\subsection*{Momentum-dependent interactions and asymptotic freedom}

A more realistic treatment starts from QCD in Coulomb
gauge~\cite{cgauge,Reinhardt:2017pyr}, where an instantaneous (yet spatially non-local) interaction arises at the Lagrangian level. This leads to momentum-dependent mass and chemical potential:
\begin{equation}
    \begin{split}
        M(p) &= \int \frac{d^3 q}{(2 \pi)^3} V(\vec{p}-\vec{q}) \frac{M(q)}{2 \tilde{E}(q)} \left( 1 - N_{\rm th}(\tilde{E}) - \bar{N}_{\rm th}(\tilde{E}) \right) \\
        \mu^\prime(p) &= \mu + \int \frac{d^3 q}{(2 \pi)^3} V(\vec{p}-\vec{q}) \frac{1}{2} \left( N_{\rm th}(\tilde{E}) - \bar{N}_{\rm th}(\tilde{E}) \right),
    \end{split}
    \label{eq:gap02}
\end{equation}
with $\tilde{E}(q) = \sqrt{q^2 + M(q)^2}$. At high densities, $M(p) \to 0$, and the vector gap equation becomes a self-consistent condition:
\begin{equation}
    \mu^\prime(p) = \mu + \mathcal{F}[\mu^\prime(p); p],
    \label{eq:gap03}
\end{equation}
where $\mathcal{F}$ is a functional of the full function $\mu^\prime(p)$. The vector density is also a functional:
\begin{equation}
    n_V[\mu^\prime(p)] = 2 N_c N_f \int \frac{d^3 q}{(2 \pi)^3} \theta(\mu^\prime(q) - q).
    \label{eq:nvf}
\end{equation}

The key point is that a Fermi momentum \( p_F \) can be defined in the dynamical case via the relation
\begin{equation}
    \mu^\prime(p_F) = p_F,
\end{equation}
which leads to the familiar result
\begin{equation}
    n_V = N_c N_f \, \frac{p_F^3}{3 \pi^2},
\end{equation}
except that \( p_F \) must now be determined self-consistently from
\begin{equation}
    p_F = \mu^\prime(p = p_F) = \mu + \mathcal{F}\left[ \mu^\prime(p); \, p = p_F \right].
    \label{eq:pFcal}
\end{equation}
Here, we make use of the trend dictated by asymptotic freedom: at large external momentum \( p = p_F \), the interaction term should drop in strength, leading to the conformal limits:
\begin{equation}
    \begin{split}
        p_F &\rightarrow \mu \\  
        n_V &\propto \mu^3 \\
        c_s^2 &\rightarrow \frac{1}{3}. 
    \end{split}
\end{equation}

Equation~\eqref{eq:pFcal} can also be used to analyze the approach toward the conformal limit. To illustrate this, we consider a simple example of a separable interaction:
\begin{equation}
    V(p, q) = V_0 \, \gamma(p) \gamma(q),
\end{equation}
with
\begin{equation}
    \gamma(p) = e^{-p^2/\Lambda^2}.
\end{equation}
It is straightforward to derive the leading-order correction to the speed of sound:
\begin{equation}
    \begin{aligned}
    c_s^2 &\approx \frac{1}{3} \left( 1 - \frac{\omega_\infty}{\mu} \,
        \left(1+\frac{2 \mu^2}{\Lambda^2}\right) \, e^{-\mu^2/\Lambda^2} \right), \\
    \omega_\infty &= G_V N_c N_f \frac{\Lambda^3}{2 \sqrt{\pi}^3}.
    \end{aligned}
\end{equation}
We provide a simple numerical illustration in Fig.~\ref{fig1}. The efficacy of the asymptotic formula (grey line) is demonstrated by comparison with the full solution obtained from direct numerical computation.

\section{Going Further}

The failure of NJL-like models to reach the conformal limit is due to their local interaction structure. By incorporating a momentum-dependent interaction consistent with asymptotic freedom, this issue is naturally resolved without arbitrary modifications.

While this study addresses high-density behavior, future work will focus on modeling confinement and deconfinement effects at intermediate densities. Theoretical models must be developed with rigor to enable meaningful comparisons to future gravitational wave data, avoiding the shortcomings of overly simplistic approaches.

\section{acknowledgments}

U. S. thanks the organizer for the kind invitation to the exciting meeting.  
P. M. L. is grateful for the fruitful discussions with J. P. Blaizot, B. Friman and W. Weise. 
He also acknowledges partial support from the Polish National Science Center (NCN) under the Opus grant no.
2022/45/B/ST2/01527.


\bibliographystyle{JHEP}
\bibliography{ref}
\end{document}